\documentclass[10pt,doublesided, doublecolumn,final]{IEEEtran}
\IEEEoverridecommandlockouts
\usepackage{float}
\usepackage{algorithm}
\usepackage[noend]{algorithmic}
\usepackage[geometry]{ifsym}
\usepackage {ifsym}
\usepackage{amssymb}
\usepackage{yhmath}
\usepackage{amscd}
\usepackage{dsfont}
\usepackage{amsmath}
\usepackage{epsfig}
\usepackage{graphics}
\usepackage{psfrag}
\usepackage{rotating}
\usepackage{amsmath}
\usepackage{amsfonts}
\usepackage{url}
\usepackage{color}
\usepackage{float}
\usepackage{balance} 
\usepackage{soul}

\newtheorem{theorem}{Theorem}
\newtheorem{definition}{Definition}

\newtheorem{corollary}{Corollary}

\newcommand{\bs}{\boldsymbol}

\newcommand{\SNR}{\mathrm{SNR}}

\newcommand{\INR}{\mathrm{INR}}

\newcommand{\sfT}{\textsf{T}}

\newcommand{\Cgicnof}{\mathcal{C}}
\newcommand{\agicnof}{\underline{\mathcal{C}}}
\newcommand{\cgicnof}{\overline{\mathcal{C}}}
\newcommand{\Ngicnof}{\mathcal{N}_{\eta}}
\newcommand{\aNgicnof}{{\underline{\mathcal{N}}_{\eta}}}
\newcommand{\cNgicnof}{\overline{\mathcal{N}}_{\eta}}

\newcommand{\overBgicnof}{\overline{\mathcal{B}}_{\mathrm{G-IC-NOF}}}

\newcommand{\GameNF}{\mathcal{G} = \left(\mathcal{K}, \left\lbrace\mathcal{A}_k \right\rbrace_{k \in \mathcal{K}},\left\lbrace
u_{k}\right\rbrace_{ k \in \mathcal{K}}\right)}

 \renewcommand{\baselinestretch}{1.02}  

\begin{document}

\title{Approximate Nash Region of the Gaussian Interference Channel with Noisy Output Feedback}
\author{Victor Quintero, Samir M. Perlaza, Jean-Marie Gorce, and H. Vincent Poor 
\thanks{Victor Quintero is with the Department of Telecommunications, Universidad del Cauca, 19003, Popay\'an, Cauca, Colombia. (vflorez@unicauca.edu.co).}
\thanks{Samir M. Perlaza and Jean-Marie Gorce are with the laboratoire CITI (a joint laboratory between the Universit\'e de Lyon, INRIA, and INSA de Lyon). 6 Avenue des Arts,  F-69621, Villeurbanne,  France. ($\lbrace$samir.perlaza, jean-marie.gorce$\rbrace$@inria.fr).}
\thanks{H. Vincent Poor and Samir M. Perlaza are with the Department of Electrical Engineering at Princeton University, Princeton, NJ 08544 USA. (poor@princeton.edu).}
\thanks{This research was supported in part by the European Commission under Marie Sk\l{}odowska-Curie Individual Fellowship No. 659316; in part by the INSA Lyon - SPIE ICS chair on the Internet of Things; in part by the Administrative Department of Science, Technology, and Innovation of Colombia (Colciencias), fellowship No. 617-2013; and in part by the U. S. National Science Foundation under Grants CNS-1702808 and ECCS-1647198.}
}
\maketitle

\begin{abstract}  
In this paper, an achievable $\eta$-Nash equilibrium ($\eta$-NE) region for the two-user Gaussian interference channel with noisy channel-output feedback is presented for all $\eta \geqslant 1$. This result is obtained in the scenario in which each transmitter-receiver pair chooses its own transmit-receive configuration in order to maximize its own individual information transmission rate. At an $\eta$-NE, any unilateral deviation by either of the pairs does not increase the corresponding individual rate by more than $\eta$ bits per channel use.  
\end{abstract}

\section{Introduction}

The interference channel (IC) is one of the simplest yet insightful multi-user channels in network information theory. An important class of ICs is the two-user Gaussian interference channel (GIC) in which there exist two point-to-point links subject to mutual interference and independent Gaussian noise sources. 
In this model, each output signal is a noisy version of the sum of the two transmitted signals affected by the corresponding channel gains. 
The analysis of this channel can be made considering two general scenarios: ($1$) a centralized scenario in which the entire network is controlled by a central
entity that configures both transmitter-receiver pairs; and ($2$) a decentralized scenario in which each transmitter-receiver pair autonomously configures its transmission-reception parameters. 
In the former, the fundamental limits are characterized by the capacity region, which is approximated to within a fixed number of bits in \cite{Etkin-TIT-2008} for the case without feedback; in \cite{Suh-TIT-2011} for the case with perfect channel-output feedback; and in \cite{SyQuoc-TIT-2015} and \cite{QPEG-TIT-2018} for the case with noisy channel-output feedback. %
In the latter, the fundamental limits are characterized by the $\eta$-Nash equilibrium ($\eta$-NE) region.
The $\eta$-NE of the GIC is approximated in the cases without feedback and with perfect channel-output feedback in \cite{Berry-TIT-2011} and \cite{Perlaza-TIT-2015}, respectively.

In this paper the $\eta$-NE region of the GIC is studied assuming that there exists a noisy feedback link from each receiver to its corresponding transmitter. The $\eta$-NE region is approximated by two regions for all $\eta \geqslant 1$: a region for which an equilibrium transmit-receive configuration is presented for each of the information rate pairs (an achievable region); and a region for which any information rate pair that is outside of this region cannot be an $\eta$-NE (impossibility region).    
The focus of this paper is on the achievable region. 

The results presented in this paper are a generalization of the results presented in \cite{Berry-TIT-2011} and \cite{Perlaza-TIT-2015}, and they are obtained thanks to the analysis of linear deterministic approximations in \cite{Quintero-ISIT-2017-1} and \cite{Quintero-INRIA-RR-2017}. 

\section{Decentralized Gaussian Interference Channels with Noisy Channel-Output Feedback} \label{SecProbFNM}

\begin{figure}[t!]
 \centerline{\epsfig{figure=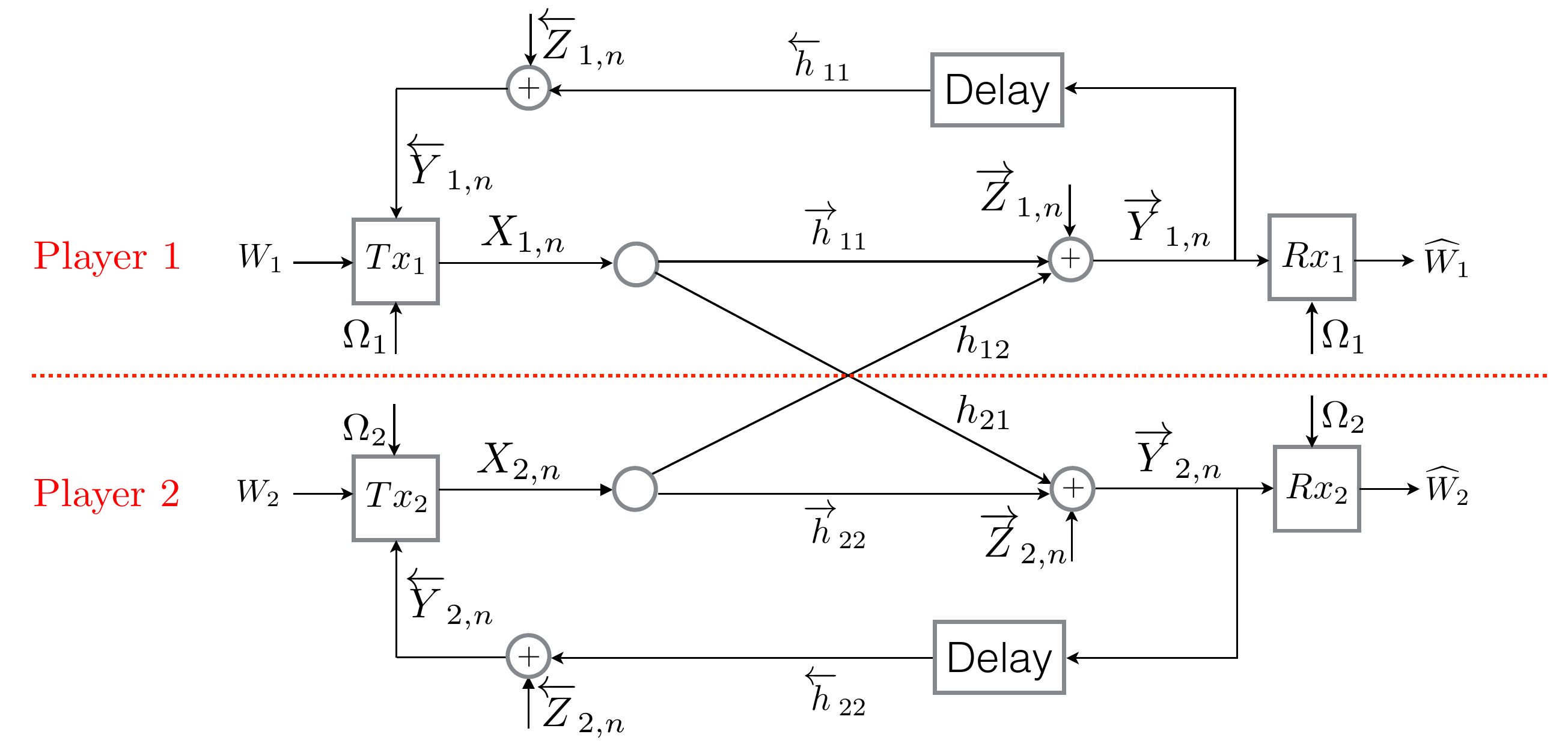,width=0.51\textwidth}}
 \caption{Two-User Decentralized Gaussian interference channel with noisy channel-output feedback at channel use~$n$.}
 \label{Fig:D-G-IC-NOF}
\end{figure}

Consider the two-user decentralized Gaussian interference channel with noisy channel-output feedback (D-GIC-NOF) depicted in Figure \ref{Fig:D-G-IC-NOF}. Transmitter $i$, with $i \in \{1,2\}$, communicates with receiver $i$ subject to the interference produced by transmitter $j$, with $j \in \{1,2\} \backslash \{i\}$. There are two independent and uniformly distributed messages, $W_i \in \mathcal{W}_i$, with $\mathcal{W}_i=\lbrace 1, 2,  \ldots, \lfloor 2^{N_iR_i} \rfloor \rbrace$, where $N_i$ denotes the fixed block-length in channel uses and $R_i$ the information transmission rate in bits per channel use. At each block, transmitter $i$  sends the codeword ${\bs{X}_{i}=\left(X_{i,1}, X_{i,2}, \ldots, X_{i,N_i}\right)^\sfT \in \mathcal{C}_i \subseteq \mathds{R}^{N_i}}$, where $\mathcal{C}_i$ is the codebook of transmitter $i$. 
The channel coefficient from transmitter $j$ to receiver $i$ is denoted by $h_{ij}$; the channel coefficient from transmitter $i$ to receiver $i$ is denoted by $\overrightarrow{h}_{ii}$; and the channel coefficient from channel-output $i$ to transmitter $i$ is denoted by $\overleftarrow{h}_{ii}$. All channel coefficients are assumed to be non-negative real numbers.
At a given channel use $n \in \{1, 2, \ldots, N\}$, with
\begin{equation}
\label{EqdefN}
N=\max(N_1,N_2),
\end{equation}
the channel output at receiver $i$ is denoted by $\overrightarrow{Y}_{i,n}$.  
During channel use $n$, the input-output relation of the channel model is given by
\begin{IEEEeqnarray}{lcl}
\label{Eqsignalyif}
\overrightarrow{Y}_{i,n}&=& \overrightarrow{h}_{ii}X_{i,n} + h_{ij}X_{j,n}+\overrightarrow{Z}_{i,n},
\end{IEEEeqnarray}
where $X_{i,n}=0$ for all $n$ such that $N \geqslant n > N_i$ and $\overrightarrow{Z}_{i,n}$ is a real Gaussian random variable with zero mean and unit variance that represents the noise at the input of receiver $i$.
Let $d>0$ be the finite feedback delay measured in channel uses. At the end of channel use $n$, transmitter $i$ observes $\overleftarrow{Y}_{i,n}$, which consists of a scaled and noisy version of $\overrightarrow{Y}_{i,n-d}$. More specifically,
\begin{IEEEeqnarray}{rcl}
\label{Eqsignalyib}
\overleftarrow{Y}_{i,n}  &=& 
\begin{cases}
 \overleftarrow{Z}_{i,n} &  \textrm{for } n \! \in \lbrace \! 1, \! 2,  \ldots, d \rbrace  \\ 
\overleftarrow{h}_{ii}\overrightarrow{Y}_{i,n-d} \! + \! \overleftarrow{Z}_{i,n}, \!  &  \textrm{for } n \! \in \lbrace  d \! + \! 1, \! d \! + \! 2, \ldots, \! N \rbrace,
 \end{cases} \quad
\end{IEEEeqnarray}
where $\overleftarrow{Z}_{i,n}$ is a real Gaussian random variable with zero mean and unit variance that represents the noise in the feedback link of transmitter-receiver pair  $i$. The random variables $\overrightarrow{Z}_{i,n}$ and $\overleftarrow{Z}_{i,n}$ are assumed to be independent.
In the following, without loss of generality, the feedback delay is assumed to be one channel use, i.e., $d=1$. 
The encoder of transmitter $i$ is defined by a set of deterministic functions $f_{i,1}^{(N)}, f_{i,2}^{(N)}, \ldots, f_{i,N_i}^{(N)}$, with $f_{i,1}^{(N)}:\mathcal{W}_i\times \mathds{N} \rightarrow \mathcal{X}_i$ and for all $n \in \lbrace 2, 3, \ldots, N_i\rbrace$, $f_{i,n}^{(N)}:\mathcal{W}_i \times \mathds{N} \times\mathds{R}^{n-1} \rightarrow \mathcal{X}_i$, such that
\begin{subequations}
\label{Eqencod}
\begin{IEEEeqnarray}{lcl}
\label{Eqencodi1}
X_{i,1}&=&f_{i,1}^{(N)}\left(W_i, \Omega_i\right),  \mbox{ and }  \\
\label{Eqencodit}
X_{i,n}&=&f_{i,n}^{(N)}\left(W_i, \Omega_i, \overleftarrow{Y}_{i,1}, \overleftarrow{Y}_{i,2}, \ldots,\overleftarrow{Y}_{i,n-1}\right),
\end{IEEEeqnarray}
\end{subequations}
where $\Omega_i$ is an additional index randomly generated. The index $\Omega_i$ is assumed to be known by both transmitter $i$ and receiver $i$, while unknown by transmitter $j$ and receiver $j$.

The components of the input vector $\bs{X}_{i}$ are real numbers subject to an average power constraint
\begin{equation}
\label{Eqconstpow}
\frac{1}{N_i}\sum_{n=1}^{N_i}\mathbb{E}_{X_{i,n}}\left[X_{i,n}^2\right] \leq 1.
\end{equation}
The decoder of receiver $i$ is defined by a deterministic function ${\psi_i^{(N)}: \mathds{N}\times\mathds{R}^{N} \rightarrow \mathcal{W}_i}$.
At the end of the communication, receiver $i$ uses the vector $\Big(\overrightarrow{Y}_{i,1}$, $\overrightarrow{Y}_{i,2}$, $\ldots$, $\overrightarrow{Y}_{i,N}\Big)$ and the index $\Omega_i$ to obtain an estimate 
\begin{IEEEeqnarray}{rcl}
\label{Eqdecoder}
\widehat{W}_i &=& \psi_i^{(N)} \Big(\Omega_{i}, \overrightarrow{Y}_{i,1}, \overrightarrow{Y}_{i,2}, \ldots, \overrightarrow{Y}_{i,N} \Big), 
\end{IEEEeqnarray} 
A \emph{transmit-receive configuration} for transmitter-receiver pair $i$, denoted by $s_i$, can be described in terms of the block-length $N_i$, the rate $R_i$, the codebook $\mathcal{C}_i$, the encoding functions $f_{i,1}^{(N)}, f_{i,2}^{(N)}, \ldots, f_{i,N_i}^{(N)}$, and the decoding function $\psi_i^{(N)}$, etc. 
The average error probability at decoder $i$ given the configurations $s_1$ and $s_2$, denoted by $p_{i}(s_1,s_2)$, is given by
\begin{IEEEeqnarray}{rcl}
\label{EqbitErrorProb}   
p_{i}(s_1,s_2)   &=& \Pr\left[ W_i \neq \widehat{W}_i \right].
\end{IEEEeqnarray}

Within this context, a rate pair $(R_1,R_2) \in \mathds{R}_+^{2}$ is said to be achievable if it complies with the following definition. 
\begin{definition}[Achievable Rate Pairs]\label{DefAchievableRatePairs}\emph{
A rate pair $(R_1,R_2) \in \mathds{R}_+^{2}$ is achievable if there exists at least one pair of configurations $(s_1,s_2)$ such that the decoding bit error probabilities $p_{1}(s_1,s_2)$ and $p_{2}(s_1,s_2)$ can be made arbitrarily small by letting the block-lengths $N_1$ and $N_2$ grow to infinity.}
\end{definition}
The aim of transmitter $i$ is to autonomously choose its transmit-receive configuration $s_i$ in order to maximize its achievable rate $R_i$. 
Note that the rate achieved by transmitter-receiver $i$ depends on both configurations $s_1$ and $s_2$ due to mutual interference. This reveals the competitive interaction between both links in the decentralized interference channel. 
The fundamental limits of the two-user D-GIC-NOF in Figure \ref{Fig:D-G-IC-NOF} can be described by six parameters: $\overrightarrow{\SNR}_i$, $\overleftarrow{\SNR}_i$, and $\INR_{ij}$, with $i \in \{1,2\}$ and $j \in \{1,2\} \backslash \{i\}$, which are defined as follows:
\begin{IEEEeqnarray}{rcl}
\label{EqSNRifwd}
\overrightarrow{\SNR}_i &\triangleq& \overrightarrow{h}_{ii}^2, \\
\label{EqINRij}
\INR_{ij}&\triangleq& h_{ij}^2 \mbox{ and } \\
\label{EqSNRibwd}
\overleftarrow{\SNR}_i&\triangleq&\overleftarrow{h}_{ii}^2\left(\overrightarrow{h}_{ii}^2 + 2\overrightarrow{h}_{ii}h_{ij}+h_{ij}^2+1\right). \quad
\end{IEEEeqnarray}

The analysis presented in this paper focuses exclusively on the case in which $\INR_{ij} > 1$ for all $i \in \lbrace 1,2 \rbrace$ and $j \in \lbrace1,2 \rbrace \setminus \lbrace i \rbrace$. The reason for exclusively considering this case follows from the fact that when $\INR_{ij} \leqslant 1$, the transmitter-receiver pair $i$ is impaired mainly by noise instead of interference. In this case, feedback does not bring a significant rate improvement.
Denote by $\Cgicnof$ the capacity region of the two-user GIC-NOF with fixed parameters $\overrightarrow{\SNR}_{1}$, $\overrightarrow{\SNR}_{2}$, $\INR_{12}$, $\INR_{21}$, $\overleftarrow{\SNR}_{1}$, and $\overleftarrow{\SNR}_{2}$. The achievable region $\agicnof$ in \cite[Theorem $2$]{QPEG-TIT-2018} and the converse region $\cgicnof$ in \cite[Theorem$3$]{QPEG-TIT-2018} approximate the capacity region $\Cgicnof$ to within $4.4$ bits \cite{QPEG-TIT-2018}. 

\section{Game Formulation}\label{SecGameFormulation}

\begin{figure}[t!]
 \centerline{\epsfig{figure=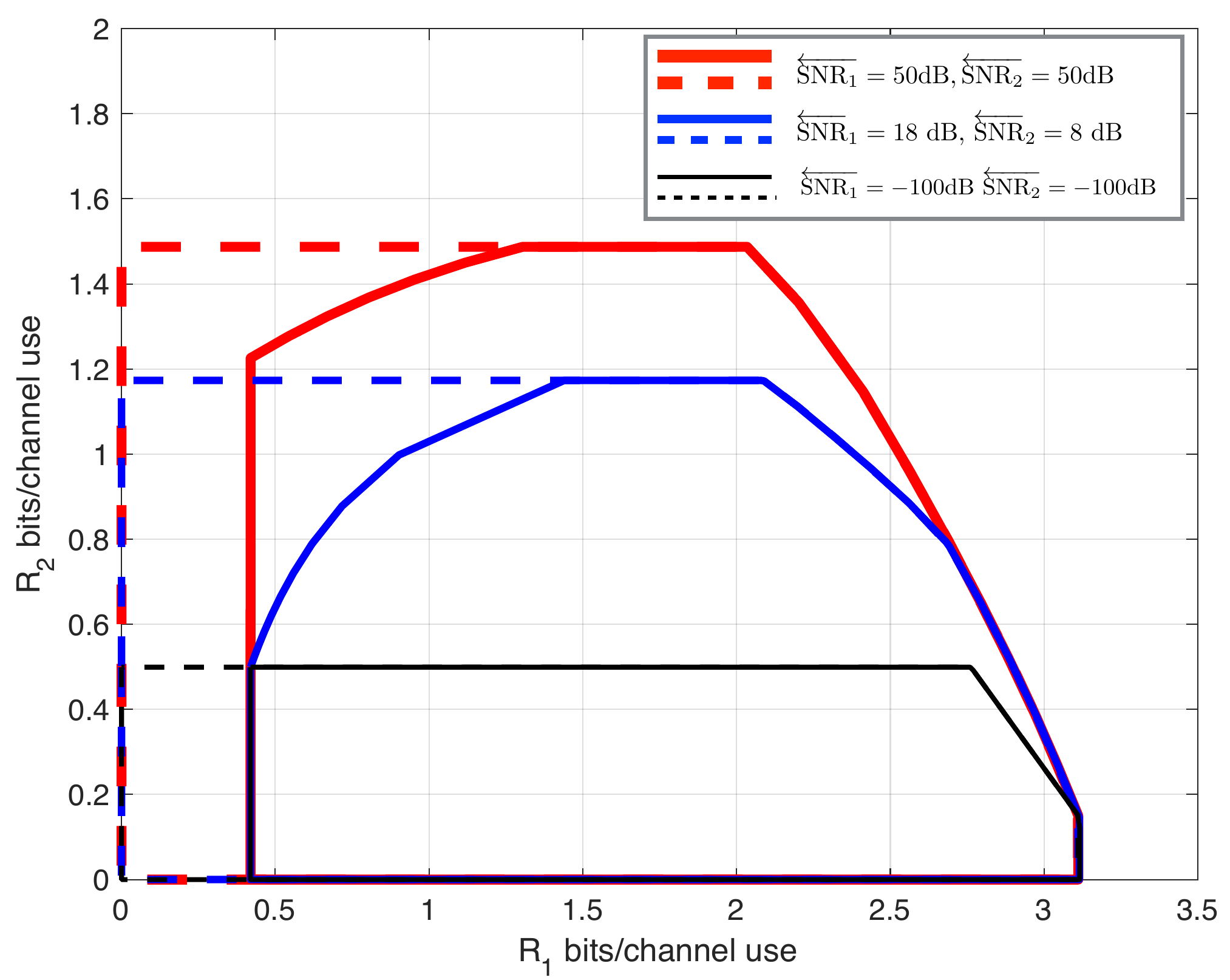,width=0.5\textwidth}}
\caption{Achievable capacity regions $\agicnof$ (dashed-lines) in \cite[Theorem $2$]{QPEG-TIT-2018} and achievable $\eta$-NE regions $\aNgicnof$ (solid lines) in Theorem \ref{TheoremMainResultGICnFB} of the two-user GIC-NOF and two-user D-GIC-NOF with parameters $\protect \overrightarrow{\SNR}_1 = 24$ dB, $\protect \overrightarrow{\SNR}_2 = 3$ dB,  $\INR_{12} = 16$ dB, $\INR_{21} = 9$ dB, $\protect \overleftarrow{\SNR}_1 \in \lbrace -100, 18, 50 \rbrace$ dB, $\protect \overleftarrow{\SNR}_2   \in \lbrace -100, 8, 50\rbrace $ dB and $\eta=1$.}  
\label{FigExamples2Z}
\end{figure}
The competitive interaction between the two transmitter-receiver pairs in the interference channel can be modeled by the following game in normal-form:
\begin{equation}\label{EqGame}
\GameNF.
\end{equation}
The set $\mathcal{K} = \lbrace 1, 2 \rbrace$ is the set of players, that is, the set of transmitter-receiver pairs. The sets $\mathcal{A}_1$ and $\mathcal{A}_2$ are the sets of actions of players $1$ and $2$, respectively. An action of a player $i \in \mathcal{K}$, which is denoted by $s_i \in \mathcal{A}_i$, is basically its transmit-receive configuration as described above. 
The utility function of player $i$ is $u_i: \mathcal{A}_1 \times \mathcal{A}_2 \rightarrow \mathds{R}_+$ and it is defined  as the information rate of transmitter $i$,
\begin{equation}
\label{EqUtility}
u_i(s_1,s_2) = \left\lbrace
\begin{array}{lcl}
R_i, & \mbox{if} &  p_{i}(s_1,s_2)  < \epsilon \\
0, & &  \mbox{otherwise,}
\end{array}
\right.
\end{equation}
where $\epsilon > 0$ is an arbitrarily small number. This game formulation for the case without feedback was first proposed in \cite{Yates-ISIT-2008} and \cite{Berry-ISIT-2008}.

A class of transmit-receive configurations that are particularly important in the analysis of this game is referred to as the set of $\eta$-Nash equilibria ($\eta$-NE), with $\eta > 0$. This type of configurations satisfy the following definition. 
\begin{definition}[$\eta$-Nash equilibrium] \label{DefEtaNE} \emph{
Given a positive real $\eta$, an action profile  $(s_1^*, s_2^*)$ is an $\eta$-Nash equilibrium (NE) in the game ${\GameNF}$,  if for all $i \in \mathcal{K}$ and for all $s_i \in \mathcal{A}_i$, it follows that
\begin{equation}\label{EqNashEquilibrium}
u_i (s_i , s_j^*) \leqslant u_i (s_i^*, s_j^*) + \eta.
\end{equation}
}
\end{definition}

Let $(s_1^*, s_2^*)$ be an $\eta$-Nash equilibrium action profile. Then, none of the transmitters can increase its own transmission rate more than $\eta$ bits per channel use by changing its own transmit-receive configuration and keeping the average bit error probability arbitrarily close to zero. 
Note that for $\eta$ sufficiently large, from Definition  \ref{DefEtaNE}, any pair of configurations can be an $\eta$-NE. Alternatively, for $\eta=0$, the definition of Nash equilibrium is obtained \cite{Nash-PNAS-1950}. In this case, if a pair of configurations is a Nash equilibrium ($\eta=0$), then each individual configuration is optimal with respect to each other. Hence, the interest is to describe the set of all possible $\eta$-NE rate pairs $(R_1,R_2)$ of the game in \eqref{EqGame} with the smallest $\eta$ for which there exists at least one equilibrium configuration pair.

The set of rate pairs that can be achieved at an $\eta$-NE is known as the $\eta$-Nash equilibrium ($\eta$-NE) region.
\begin{definition}[$\eta$-NE Region] \label{DefNERegion} \emph{
Let $\eta > 0$ be fixed. An achievable rate pair $(R_1,R_2)$ is said to be in the $\eta$-NE region of the game $\GameNF$ if there exists a pair $(s_1^*, s_2^*) \in \mathcal{A}_1 \times \mathcal{A}_2$  that is  an  $\eta$-NE and the following holds:
\begin{eqnarray}
u_1 (s_1^* , s_2^*)  =  R_1 & \mbox{ and } & u_2 (s_1^* , s_2^*)  =  R_2. 
\end{eqnarray}
}
\end{definition}
The $\eta$-NE regions of the two-user GIC with and without perfect channel-output feedback have been approximated to within a constant number of bits in \cite{Berry-TIT-2011} and \cite{Perlaza-TIT-2015}, respectively.  The next section introduces a generalization of these results.

\section{Main Results}\label{SectMainResultsDGICNOF}
\subsection{Achievable $\eta$-Nash Equilibrium Region} \label{SectMainResultsDGICNOFAchiev}
Let the $\eta$-NE region (Definition \ref{DefNERegion}) of the D-GIC-NOF be denoted by $\Ngicnof$. 
This section introduces a region $\aNgicnof \subseteq \Ngicnof$ that is achievable using a coding scheme that combines rate splitting \cite{Han-TIT-1981}, common randomness \cite{Berry-TIT-2011, Perlaza-TIT-2015}, block Markov superposition coding \cite{Cover-TIT-1981} and backward decoding \cite{Willems-PhD-1982}. In the following, this coding scheme is referred to as randomized Han-Kobayashi scheme with noisy channel-output feedback (RHK-NOF). This coding scheme is presented in \cite{Quintero-INRIA-RR-2017} and uses the same techniques of the schemes in \cite{Berry-TIT-2011} and \cite{Perlaza-TIT-2015}. Therefore, the focus of this section is on the results rather than the description of the scheme. A motivated reader is referred to \cite{Quintero-Thesis-2017}.  
The RHK-NOF is proved to be an $\eta$-NE action profile with $\eta \geqslant 1$. That is, any unilateral deviation from the RHK-NOF by any of the transmitter-receiver pairs might lead to an individual rate improvement which is at most one bit per channel use. 
\begin{figure}[t!]
 \centerline{\epsfig{figure=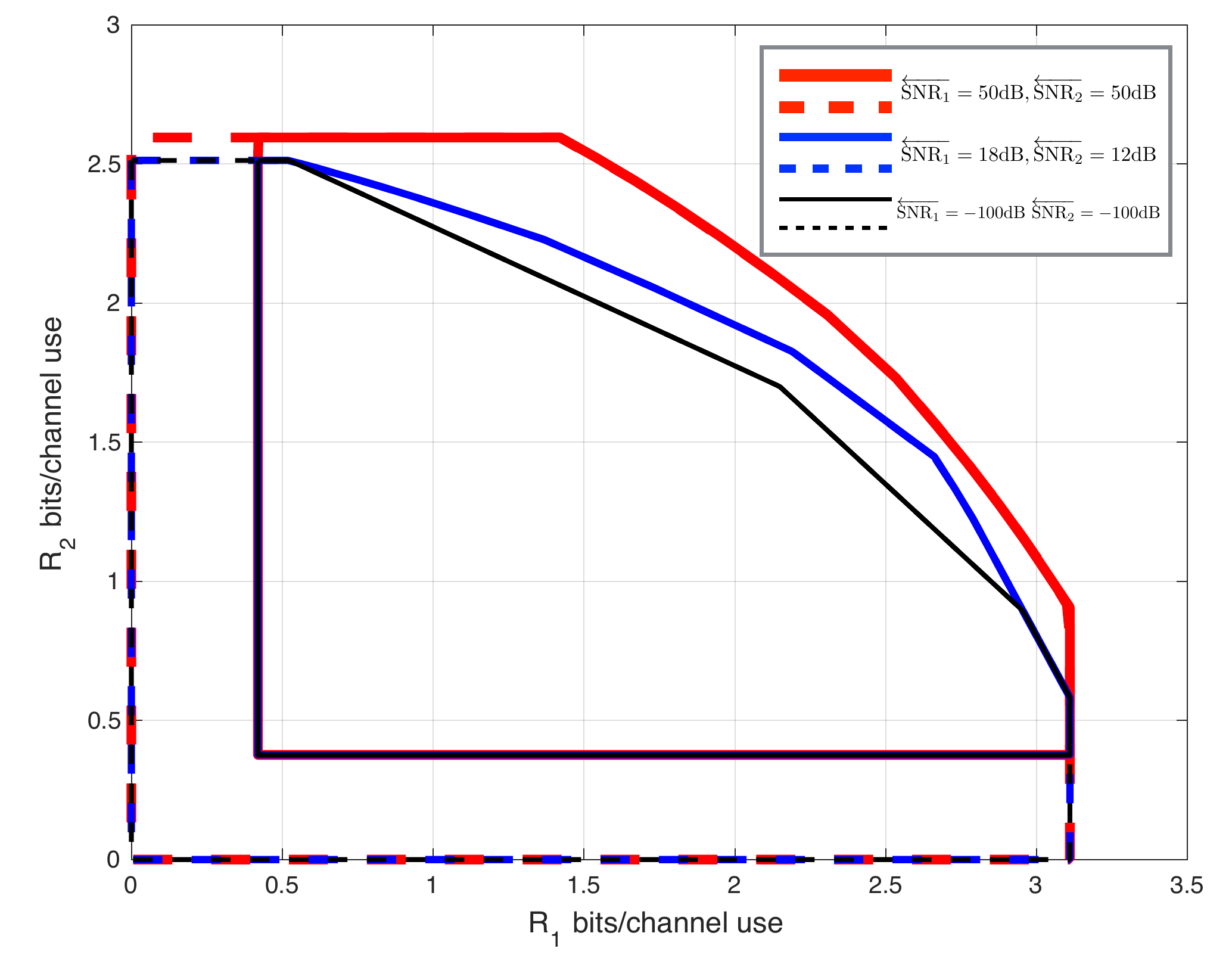,width=0.52\textwidth}}
\caption{Achievable capacity regions $\agicnof$ (dashed-lines) in \cite[Theorem $2$]{QPEG-TIT-2018} and achievable $\eta$-NE regions $\aNgicnof$ (solid lines) in Theorem \ref{TheoremMainResultGICnFB} of the two-user GIC-NOF and two-user D-GIC-NOF with parameters $\protect \overrightarrow{\SNR}_1 = 24$ dB, ${\protect \overrightarrow{\SNR}_2 = 18}$ dB,  $\INR_{12} = 16$ dB, $\INR_{21} = 10$ dB, $\protect \overleftarrow{\SNR}_1 \in \lbrace -100, 18, 50 \rbrace$ dB, $\protect \overleftarrow{\SNR}_2   \in \lbrace -100, 12, 50\rbrace $ dB and $\eta=1$.}  
  \label{FigExamples1}
\end{figure}
The description of the achievable $\eta$-Nash region $\aNgicnof$ is presented using the constants $a_{1,i}$; the functions $a_{2,i}:[0,1] \rightarrow \mathds{R}_{+}$,  $a_{l,i}:[0,1]^2\rightarrow \mathds{R}_{+}$, with $l \in \lbrace 3, \ldots, 6 \rbrace$; and $a_{7,i}:[0,1]^3\rightarrow \mathds{R}_{+}$, which are defined as follows, for all $i \in \lbrace 1, 2 \rbrace$, with $j \in \lbrace 1, 2 \rbrace \setminus \lbrace i \rbrace$:
\begin{subequations}
\label{Eq-a}
\begin{IEEEeqnarray}{rcl}
\label{Eq-a1}
a_{1,i}  &=&  \frac{1}{2}\log \left(2+\frac{\overrightarrow{\SNR_{i}}}{\INR_{ji}}\right)-\frac{1}{2}, \\
\label{Eq-a2}
a_{2,i}(\rho) &=& \frac{1}{2}\log \Big(b_{1,i}(\rho)+1\Big)-\frac{1}{2}, \\
\nonumber
a_{3,i}(\rho,\mu) &=& \!  \frac{1}{2}\!\log\! \left(\! \frac{\! \overleftarrow{\SNR}_i \! \Big(b_{2,i}(\rho)+2\Big)+b_{1,i}(1)+1}{\overleftarrow{\SNR}_i\Big( \! \left(1\!-\!\mu\right) \! b_{2,i}( \rho )\!+\!2\Big)\!+\! b_{1,i}( 1  ) \!+ \!1} \! \right) \!, \\
\label{Eq-a3}
\end{IEEEeqnarray}
\begin{IEEEeqnarray}{rcl}
\label{Eq-a4}
a_{4,i}(\rho,\mu) &=& \frac{1}{2}\log \bigg(\Big(1-\mu\Big)b_{2,i}(\rho)+2 \bigg)-\frac{1}{2}, \\
\nonumber
a_{5,i}(\rho,\mu) &=& \frac{1}{2}\log \left(2+\frac{\overrightarrow{\SNR}_{i}}{\INR_{ji}}+\Big(1-\mu\Big)b_{2,i}(\rho)\right)-\frac{1}{2},\\
\label{Eq-a5} \\
\nonumber
a_{6,i}(\rho,\mu) &=& \frac{1}{2}\!\log\! \left(\!\frac{\overrightarrow{\SNR}_{i}}{\INR_{ji}}\bigg(\Big(1\!-\!\mu\Big)b_{2,j}(\rho)\!+\!1\bigg)\!+\!2\right)\!-\!\frac{1}{2}, 
\label{Eq-a6}\\
\nonumber
a_{7,i}(\rho,\!\mu_1\!,\!\mu_2\!) &=& \frac{1}{2}\!\log \Bigg(\!\frac{\overrightarrow{\SNR}_{i}}{\INR_{ji}}\bigg(\Big(1\!-\!\mu_i\Big)b_{2,j}(\rho)\!+\!1\bigg) \\
\label{Eq-a7}
& &  +\Big(1\!-\!\mu_j\Big)b_{2,i}(\rho)+2\Bigg)\!-\!\frac{1}{2},
\end{IEEEeqnarray}
\end{subequations}
where the functions $b_{l,i}:[0,1]\rightarrow \mathds{R}_{+}$, with $l \in \lbrace1, 2 \rbrace$ are defined as follows: 
\begin{subequations}
\label{Eqfnts}
\begin{IEEEeqnarray}{rcl}
\label{Eqb1i}
b_{1,i}(\rho)&=&\overrightarrow{\SNR}_{i}+2\rho\sqrt{\overrightarrow{\SNR}_{i}\INR_{ij}}+\INR_{ij} \mbox{ and } \\
\label{Eqb5i}
b_{2,i}(\rho)&=&\Big(1-\rho\Big)\INR_{ij}-1.
\end{IEEEeqnarray}
\end{subequations}
Note that the functions in \eqref{Eq-a} and \eqref{Eqfnts} depend on $\overrightarrow{\SNR}_{1}$, $\overrightarrow{\SNR}_{2}$, $\INR_{12}$, $\INR_{21}$, $\overleftarrow{\SNR}_{1}$, and $\overleftarrow{\SNR}_{2}$, however as these parameters are fixed in this analysis, this dependence is not emphasized in the definition of these functions. Finally, using this notation, the achievable $\eta$-NE region is presented by Theorem~\ref{TheoremMainResultGICnFB} on the next page. 
The proof of Theorem \ref{TheoremMainResultGICnFB} is presented in \cite{Quintero-INRIA-RR-2017}.
The inequalities in \eqref{EqNashGAchievableregion} are additional conditions to those defining the region $\agicnof$ in \cite[Theorem 2]{QPEG-TIT-2018}. More specifically, the $\eta$-NE region is described by the intersection of the achievable region $\agicnof$ and the set of rate pairs $(R_1,R_2)$ satisfying  \eqref{EqNashGAchievableregion}.
\begin{figure}[t!]
 \centerline{\epsfig{figure=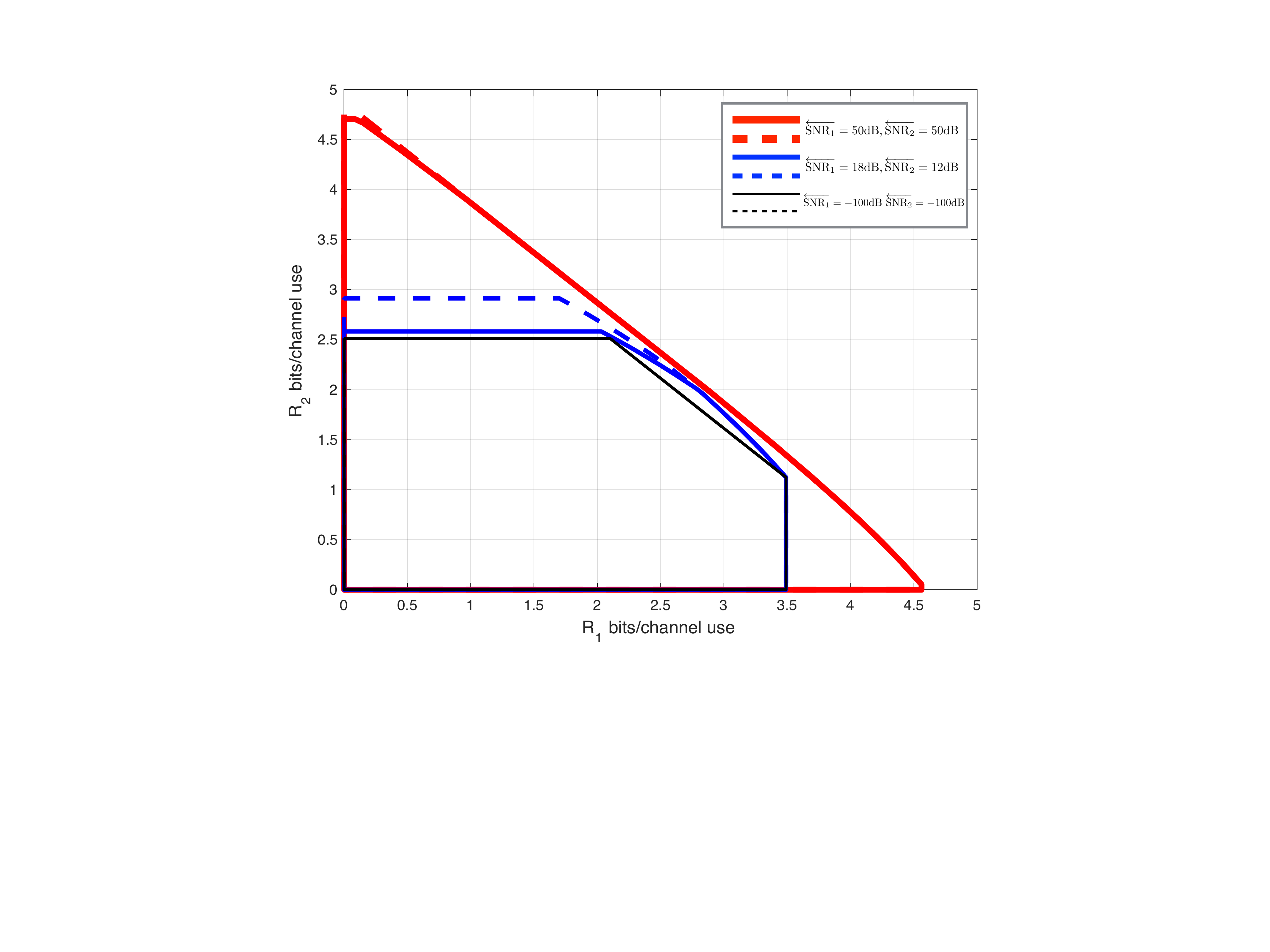,width=0.55\textwidth}}
\caption{Achievable capacity regions $\agicnof$ (dashed-lines) in \cite[Theorem $2$]{QPEG-TIT-2018} and achievable $\eta$-NE regions $\aNgicnof$ (solid lines) in Theorem \ref{TheoremMainResultGICnFB} of the two-user GIC-NOF and two-user D-GIC-NOF with parameters $\protect \overrightarrow{\SNR}_1 = 24$ dB, ${\protect \overrightarrow{\SNR}_2 = 18}$ dB,  $\INR_{12} = 48$ dB, $\INR_{21} = 30$ dB, $\protect \overleftarrow{\SNR}_1 \in \lbrace -100, 18, 50 \rbrace$ dB, $\protect \overleftarrow{\SNR}_2   \in \lbrace -100, 12, 50\rbrace $ dB and $\eta=1$.}  
 \label{FigExamples2}
\end{figure}

\begin{figure*}[t!]
\begin{theorem}\label{TheoremMainResultGICnFB}\emph{Let $\eta \geqslant 1$ be fixed. The achievable $\eta$-NE region $\aNgicnof$ is given by the closure of all possible achievable rate pairs $(R_1,R_2) \in \agicnof$ in \cite[Theorem $2$]{QPEG-TIT-2018} that  satisfy, for all $i \in \lbrace 1,2 \rbrace$ and $j \in \lbrace1,2 \rbrace \setminus \lbrace i \rbrace$, the following conditions:
\begin{subequations}
\label{EqNashGAchievableregion}
\begin{IEEEeqnarray}{rcl}
\label{EqNashGAchievableregion1}
R_i  & \geqslant &  \Big(a_{2,i}(\rho) -a_{3,i}(\rho,\mu_j)-a_{4,i}(\rho,\mu_j)-\eta\Big)^+,\\
\label{EqNashGAchievableregion2}
R_i &\leqslant& \min\Big( a_{2,i}(\rho)+a_{3,j}(\rho,\mu_i)+a_{5,j}(\rho,\mu_i)-a_{2,j}(\rho)+\eta,  \\
\nonumber
& &  a_{3,i}(\rho,\mu_j)+a_{7,i}(\rho,\mu_1,\mu_2)+2a_{3,j}(\rho,\mu_i)+a_{5,j}(\rho,\mu_i)-a_{2,j}(\rho)+\eta,   \\
\nonumber
& &  \! a_{2,i}(\rho) \! + \! a_{3,i}(\rho,\mu_j) \! + \!  2a_{3,j}(\rho,\mu_i) \! + \!  a_{5,j}(\rho,\mu_i) \!  + \!  a_{7,j}(\rho,\mu_1,\mu_2) \!  - \!  2a_{2,j}(\rho) + 2\eta \Big),  \\
\label{EqNashGAchievableregion3}
R_1+R_2 & \leqslant & a_{1,i} \!  + \!   a_{3,i}(\rho,\mu_j)\! +\! a_{7,i}(\rho,\mu_1,\mu_2)\! +\! a_{2,j}(\rho)\! +\! a_{3,j}(\rho,\mu_1)\! -\! a_{2,i}(\rho) + \eta, 
\end{IEEEeqnarray}
\end{subequations}
for all $\left(\rho, \mu_1, \mu_2\right) \in \left[0,\left(1-\max\left(\frac{1}{\INR_{12}},\frac{1}{\INR_{21}}\right) \right)^+\right]\times[0,1]\times[0,1]$.
}
\end{theorem}
\noindent\rule{18cm}{0.6pt}
\vspace{-5mm}
\end{figure*}

Figure \ref{FigExamples2Z} shows the achievable region $\agicnof$ in \cite[Theorem $2$]{QPEG-TIT-2018} of a two-user centralized GIC-NOF and the achievable $\eta$-NE region  $\aNgicnof$ in Theorem \ref{TheoremMainResultGICnFB} of a two-user D-GIC-NOF with parameters $\overrightarrow{\SNR}_1 = 24$ dB, $\overrightarrow{\SNR}_2 = 3$ dB,  $\INR_{12} =16 $ dB, $\INR_{21} = 9$ dB, $\overleftarrow{\SNR}_1 \in \lbrace -100, 18, 50 \rbrace$ dB, $\overleftarrow{\SNR}_2   \in \lbrace -100, 8, 50\rbrace $ dB and $\eta=1$.  
Note that in this case, the feedback parameter $\overleftarrow{\SNR_{2}}$ does not have an effect on the achievable $\eta$-NE region $\aNgicnof$ and the achievable capacity region $\agicnof$  (\cite[Theorem $2$]{QPEG-TIT-2018}). This is due to the fact that when one transmitter-receiver pair is in low interference regime (LIR) and the other transmitter-receiver pair is in high interference regime (HIR), feedback is useless on the transmitter-receiver pair in HIR \cite{Quintero-Thesis-2017, QPEG-TC-2018}.

Figure \ref{FigExamples1} shows the achievable region $\agicnof$ in \cite[Theorem $2$]{QPEG-TIT-2018} of a two-user centralized GIC-NOF and the achievable $\eta$-NE region  $\aNgicnof$ in Theorem \ref{TheoremMainResultGICnFB} of a two-user D-GIC-NOF with parameters $\overrightarrow{\SNR}_1 = 24$ dB, $\overrightarrow{\SNR}_2 = 18$ dB,  $\INR_{12} =16 $ dB, $\INR_{21} = 10$ dB, $\overleftarrow{\SNR}_1 \in \lbrace -100, 18, 50 \rbrace$ dB, $\overleftarrow{\SNR}_2   \in \lbrace -100, 12, 50\rbrace $ dB and $\eta=1$.  
Figure \ref{FigExamples2} shows the achievable region $\agicnof$ in \cite[Theorem $2$]{QPEG-TIT-2018} of a two-user centralized GIC-NOF and the achievable $\eta$-NE region  $\aNgicnof$ in Theorem \ref{TheoremMainResultGICnFB} of a two-user D-GIC-NOF with parameters $\overrightarrow{\SNR}_1 = 24$ dB, $\overrightarrow{\SNR}_2 = 18$ dB,  $\INR_{12} =48$ dB, $\INR_{21} = 30$ dB, $\overleftarrow{\SNR}_1 \in \lbrace -100, 18, 50 \rbrace$ dB, $\overleftarrow{\SNR}_2   \in \lbrace -100, 12, 50\rbrace $ dB and $\eta=1$.  
In this case, the achievable $\eta$-NE region $\aNgicnof$ in Theorem \ref{TheoremMainResultGICnFB} and achievable region $\agicnof$ on the capacity region \cite[Theorem 2]{QPEG-TIT-2018} are almost identical, which implies that in the cases in which $\overrightarrow{\SNR_{i}} < \INR_{ij}$, for both $i \in \lbrace 1, 2 \rbrace$, with $j \in \lbrace 1, 2 \rbrace \setminus \lbrace i \rbrace$, the achievable $\eta$-NE region is almost the same as the achievable capacity region in the centralized case studied in \cite{QPEG-TIT-2018}. 
At low values of $\overleftarrow{\SNR}_1$ and $\overleftarrow{\SNR}_2$, the achievable $\eta$-NE region approaches the rectangular region reported in \cite{Berry-TIT-2011} for the case of the two-user decentralized GIC (D-GIC). Alternatively, for high values of $\overleftarrow{\SNR}_1$ and $\overleftarrow{\SNR}_2$, the achievable $\eta$-NE region approaches the region reported in \cite{Perlaza-TIT-2015} for the case of the two-user decentralized GIC with perfect channel-output feedback (D-GIC-POF). These observations are formalized by the following corollaries. 

Denote by $\aNgicnof_{\mathrm{PF}}$ the achievable $\eta$-NE region of the two-user D-GIC-POF presented in \cite{Perlaza-TIT-2015}. The region $\aNgicnof_{\mathrm{PF}}$ can be obtained as a special case of Theorem \ref{TheoremMainResultGICnFB} as shown by the following corollary. 

\begin{corollary}[$\eta$-NE Region with Perfect Output Feedback] \emph{Let $\aNgicnof_{\mathrm{PF}}$ denote the achievable $\eta$-NE region of the two-user D-GIC-POF with fixed parameters $\overrightarrow{\SNR}_i$ and $\INR_{ij}$, with $i \in \{1,2\}$ and $j \in \{1,2\} \backslash \{i\}$. Then, the following holds:
\begin{IEEEeqnarray}{l}
\aNgicnof_{\mathrm{PF}} = \hspace{-2mm}\lim_{
\tiny
\begin{array}{l}
\overleftarrow{\SNR}_1\rightarrow \infty\\  
\overleftarrow{\SNR}_2 \rightarrow \infty
\end{array}}
\hspace{-5mm}\aNgicnof.
\end{IEEEeqnarray}
}
\end{corollary}
Denote by $\aNgicnof_{\mathrm{WF}}$ the achievable $\eta$-NE region of the two-user D-GIC presented in \cite{Berry-TIT-2011}. The region $\aNgicnof_{\mathrm{WF}}$ can be obtained as a special case of Theorem \ref{TheoremMainResultGICnFB} as shown by the following corollary. 

\begin{corollary}[$\eta$-NE Region without Output Feedback ] \emph{Let $\aNgicnof_{\mathrm{WF}}$ denote the achievable $\eta$-NE region of the two-user D-GIC, with fixed parameters $\overrightarrow{\SNR}_i$ and $\INR_{ij}$, with $i \in \{1,2\}$ and $j \in \{1,2\} \backslash \{i\}$. Then, the following holds:
\begin{IEEEeqnarray}{l}
\aNgicnof_{\mathrm{WF}} = \hspace{-2mm}\lim_{
\tiny
\begin{array}{r}
\overleftarrow{\SNR}_1 \rightarrow 0 \\
\overleftarrow{\SNR}_2  \rightarrow 0 \\
\rho =  0.
\end{array}}
\hspace{-5mm}\aNgicnof.
\end{IEEEeqnarray}
}
\end{corollary}

\subsection{Imposibility Region} \label{SectMainResultsDCGICNOF}

This section introduces an imposibility region, denoted by $\cNgicnof$. That is, $\cNgicnof \supseteq \Ngicnof$. More specifically, any rate pair $\left(R_1,R_2\right) \in \cNgicnof^{\mathsf{c}}$ is not an $\eta$-NE. This region is described in terms of the convex region $\overBgicnof$. Here, for the case of the two-user D-GIC-NOF, the region $\overBgicnof$ is given by the closure of the rate pairs $(R_1,R_2) \in \mathds{R}_{+}^2$ that satisfy for all $i \in \lbrace 1, 2 \rbrace$, with $j\in\lbrace 1, 2 \rbrace\setminus\lbrace i \rbrace$:

\begin{IEEEeqnarray}{lcl}
\nonumber
\overBgicnof & = & \Big \lbrace (R_1,R_2)\in \mathds{R}_{+}^2 : R_i \geqslant L_i, \\
\label{EqBiCGaussian}
& & \mbox{ for all } i \in \mathcal{K}=\lbrace 1, 2 \rbrace \Big \rbrace, 
\end{IEEEeqnarray}
where, 
\begin{IEEEeqnarray}{lcl}
\label{EqLiGaussian}
L_i & \triangleq &   \left(\frac{1}{2}\log\left( 1 + \frac{\overrightarrow{\SNR}_i}{1 + \INR_{ij}}\right)-\eta\right)^+.
\end{IEEEeqnarray}
Note that $L_i$ is the rate achieved by the transmitter-receiver pair $i$ when it saturates the power constraint in \eqref{Eqconstpow} and treats interference as noise. Following this notation, the imposibility region of the two-user GIC-NOF, i.e., $\cNgicnof$, can be described as follows.

\begin{theorem}\label{TheoremMainResultNCGICnFB}\emph{
Let $\eta \geqslant 1$ be fixed. The imposibility region $\cNgicnof$ of the two-user D-GIC-NOF  is given by the closure of all possible non-negative rate pairs $(R_1,R_2) \in \cgicnof \cap \overBgicnof$ for all $\rho \in [0,1]$. 
}
\end{theorem}

The impossibility region in Theorem \ref{TheoremMainResultNCGICnFB} has been first presented in \cite{Perlaza-TIT-2015} and it is very loose in this case. A better impossibility region is presented in \cite{Quintero-Thesis-2017}.

\section{Conclusions}\label{SecConclusions}
In this paper, an achievable $\eta$-Nash equilibrium ($\eta$-NE) region for the two-user Gaussian interference channel with noisy channel-output feedback has been presented for all $\eta \geqslant 1$. 
This result generalizes the existing achievable regions of the $\eta$-NE for the the cases without feedback and with perfect channel-output feedback.

\balance
\bibliographystyle{IEEEtran}
\bibliography{IT-GT}
\balance

\end{document}